\documentclass[fleqn,12pt]{article}
\usepackage{latexsym}
\usepackage[usenames]{color}
\usepackage{amssymb}
\usepackage{times}
\usepackage{graphicx}
\usepackage{caption}
\usepackage{amsmath,amsfonts,amsthm}
\usepackage{graphicx}
\usepackage{setspace}
\usepackage{pdfpages}
\usepackage{enumitem}
\usepackage{indentfirst}
\usepackage[left=2cm,top=2cm,right=2cm,footskip=20pt, bottom = 2cm,nohead]{geometry}
\usepackage{bbm}
\usepackage{nameref}
\usepackage{subfigure}
\usepackage{authblk}
\usepackage{helvet}
\usepackage{url}
\usepackage{natbib}

\let\oldref\ref
\renewcommand{\ref}[1]{(\oldref{#1})}

\newcommand{\T}{^{\ensuremath{\mathsf{T}}}}           
\newcommand{\mrsid}{{\sc \texttt{Mr}.~\texttt{Sid}}}

\newcommand{\argmin}{\operatornamewithlimits{argmin}}
\newcommand{\diag}{\operatornamewithlimits{diag}}

\providecommand{\mb}[1]{\boldsymbol{#1}}
\newcommand{\bx}{\mb{x}}
\newcommand{\by}{\mb{y}}
\newcommand{\bX}{\mb{X}}
\newcommand{\bY}{\mb{Y}}

\begin{document}

\title{An M-Estimator for Reduced-Rank High-Dimensional Linear Dynamical System Identification
}
\author[a]{Shaojie Chen}
\author[b]{Kai Liu}
\author[c]{Yuguang Yang}
\author[a]{Yuting Xu}
\author[d]{Seonjoo Lee}
\author[a]{Martin Lindquist}
\author[a]{Brian S. Caffo}
\author[e,f]{Joshua T. Vogelstein}
\affil[a]{\small Dept. of Biostatistics, Johns Hopkins Bloomberg School of Public Health}
\affil[b]{Dept. of Neuroscience, Johns Hopkins University}
\affil[c]{Dept. of Chemical and Biomolecular Engineering, Johns Hopkins University}
\affil[d]{Dept. of Psychiatry and Department of Biostatistics, Columbia University}
\affil[e]{Child Mind Institute}
\affil[f]{Dept. of Biomedical Enginnering and Institute for Computational Medicine, Johns Hopkins University}
\date{}
\maketitle
\section*{Abstract}
High-dimensional time-series data are becoming increasingly abundant across a wide variety of domains, spanning economics, neuroscience,  particle physics, and cosmology.  Fitting statistical models to such data, to enable parameter estimation and time-series prediction, is an important computational primitive.
Existing methods, however, are unable to cope with the high-dimensional nature of these problems, due to both computational and statistical reasons.  We mitigate both kinds of issues via proposing an M-estimator for Reduced-rank System IDentification (\mrsid). A combination of low-rank approximations, $\ell_1$ and $\ell_2$ penalties, and some numerical linear algebra tricks, yields an estimator that is computationally efficient and numerically stable.  Simulations and real data examples demonstrate the utility of this approach in a variety of problems.  In particular, we demonstrate that \mrsid~can estimate spatial filters, connectivity graphs, and time-courses from native resolution functional magnetic resonance imaging data.  Other applications and extensions are immediately available, as our approach is a generalization of the classical Kalman Filter-Smoother Expectation-Maximization algorithm.
\\
\textbf{\emph{keywords}: state-space model, parameter estimation, sparsity, imaging processing, fMRI}
\section{Introduction}

High-dimensional time-series data are becoming increasingly abundant across a wide variety of domains, spanning economics \citep{Johansen1988}, neuroscience \citep{Friston2003a},   and cosmology \citep{Xie2013a}.
Fitting statistical models to such data, to enable parameter estimation and time-series prediction, is an important computational primitive.
Linear dynamical systems (LDS) models are amongst the most popular and powerful, because of their intuitive nature and ease of implementation \citep{Kalman1963}.   The famous Kalman Filter Smoother is one of the most popular and powerful methods for time-series prediction in an LDS, given known parameters \citep{Kalman1960a}.
In practive, however, for many LDSs, the parameters are unknown, and must be estimated, a process often called \emph{system identification} in this domain \citep{Ljung1998}.  To date, there does not exist, to our knowledge, a methodology that provides parameter estimates and predictions from ultrahigh-dimensional time-series data (for example, $p > 10$,$000$).

The challenges associated high-dimensional time-series estimation and prediction are multifold.  First, na\"{i}vely such models include dense $p \times p$ matrices, which often are too large even to store, and much too large to invert in memory.  For large sparse matrices, recently, several efforts to invert them using a series of computational tricks are promising, though still extremely computationally expensive  \citep{Hsieh2013, Banerjee2013a}.
Second, estimators behave poorly, due to numerical instability issues.
Reduced rank LDS models reduce the number of latent dimensions, and therefore partially address this problem \citep{CHEN1989}.  However, without further constraints, the dimensionality of the latent states must be so small as to significantly decrease the predictive capacity of the resulting model.  Third, even after addressing these problems, the time to compute all the necessary quantities can be overly burdensome. Distributed memory implementations, such as using Spark, might be above to overcome this problem, but it would at additional costs and set-up burden, as it would require a Spark cluster \citep{Zaharia2010}.

We address all three of these issues with our M-estimator for Reduced-rank  System IDentification (\mrsid).  By assuming the dimensionality of the latent state space is small (reduced rank), relative to the ambient, or observed, space dimensionality, we can significantly improve computational tractability and estimation accuracy. By further penalizing the estimators, with $\ell_1$ and/or $\ell_2$ penalties, utilizing prior knowledge on the structure of the parameters, we gain further estimation accuracy in this high-dimensional but relatively low-sample size regime.  Finally, by employing several numerical linear algebra tricks, we can bring the computational burden down significantly.

These three techniques together enable us to obtain highly accurate estimates in a variety of simulation settings.  \mrsid~is, in fact, a generalization of the now classic Baum-Welch expectation maximization algorithm, commonly used for system identification in much lower dimensional linear dynamical systems \citep{rabiner1989tutorial}. We numerically show that the hyper-parameters can be selected minimizing prediction error on held-out data.  Finally, we use \mrsid~to estimate functional connectomes from motor cortex.  \mrsid~enables us to estimate the regions, rather than imposing some prior parcellation on the data, as well as estimate sparse connectivity between regions.  \mrsid~reliably estimates these connectomes, as well as predicts the held-out time-series data.  To our knowledge, this is the first time anybody has been able to estimate partitions and functional connectomes directly from the high-dimensional data, with a single unified approach.  To enable extensions, generalizations, and additional applications, the functions and code for generating each of the figures is freely available from \url{https://github.com/shachen/PLDS/} (referred to as the PLDS Git Repo in following sections).

\section{The Model}

In statistical data analysis, one often encounter some observed signals and also some unobserved or latent variables, which we denote as $\bY=(\by_1,\ldots,\by_T)$ and $\bX=(\bx_1,\ldots,\bx_T)$ respectively. By the Bayesian rule, the joint probability of $\bY$ and $\bX$ is $P(\bX,\bY)=P(\bY|\bX) P(\bX)$. The conditional distribution $P(\mb{Y}|\mb{X})$ and prior can both be represented as a product of marginals:
\begin{equation}
\begin{aligned}
P(\mb{Y}|\mb{X}) &= \prod_{t=1}^T P(\by_t | \by_{0:t-1}, x_{0:t}), \\
P(\bX) &= P(x_0) \prod_{t=1}^T P(x_t | x_{0:t-1}).
\end{aligned}
\end{equation}

The generic time-invariant state-space model makes the following simplifying assumptions
\begin{equation}
\begin{aligned}
P(\by_t | \by_{0:t-1}, x_{0:t})  &\approx P(\by_t | \bx_t), \\
P(x_t | x_{0:t-1}) &\approx P(x_t | x_{t-1}).
\end{aligned}
\end{equation}

Finally, a lineary dynamical system assumes that both of the above terms are linear Gaussian functions, which, when written as an iterative random process, yields the standard matrix update rules:
\begin{equation} \label{eq:model}
\begin{aligned}
&\bx_{t+1}=A\bx_t+\bY'_t, \quad \bY'_t\sim N(\mathbf{0},Q),\quad \bx_0 \sim N(\mathbf{\pi}_0,V_0)\\
&\by_t=C\bx_t+\mathbf{v}_t,\qquad \mathbf{v}_t\sim N(\mathbf{0},R),
\end{aligned}
\end{equation}
where $A$ is the $d\times d$ state transition matrix and $C$ is the $p \times d$ generative matrix. $\bx_t$ is a $d\times 1$ vector and $\by_t$ is a $p\times 1$ vector.
The output noise covariance $R$ is $p\times p$, while the state noise covariance $Q$ is $d\times d$. Initial state mean $\mathbf{\pi}_0$ is $d\times 1$ and covariance $V_0$ is $d \times d$.

The model can be thought as the continuous version of the hidden Markov model (HMM), where the columns of $C$ stands for the hidden states. The difference is that in this model, what one observe at each time point is not a single state, but a linear combination of multiple states. $\bx_t$ is the weights in the linear combination. $A$ matrix is the analogy of the state transition matrix, which describe how the weights $\bx_t$ evolve over time. Another difference is that LDS contains two white noise terms, which are captured by the $Q$ and $R$ matrices.

Without applying further constraints, the model itself is unidentifiable. Supplemental constraints are are thus introduced to address both identifiability and utility. Three basic constraints are required to make the model identifiable:
\vspace*{-3mm}
\begin{equation*}\label{eq:constraints1}
\begin{aligned}
&\textbf{Constraint 1: }Q \text{ is the identity matrix}\\
&\text{\textbf{Constraint 2:} the ordering of the columns of } C \text{ is fixed based on their norms}\\
&\textbf{Constraint 3: } V_0=\mathbf{0}
\end{aligned}
\end{equation*}
Note that the first two constraints follow directly from Roweis and Ghahramani (1999) \citep{roweis1999unifying}.

The logic for Constraint 1 is as follows. Since Q is a covariance matrix, it is symmetric and positive semidefinite and thus can be expressed in the form $E\Lambda E^T$ where $E$ is a rotation matrix of eigenvectors and $\Lambda$ is a diagonal matrix of eigenvalues. Thus, for any model where $Q$ is not the identity matrix, one can generate an equivalent model using a new state vector $\bx^{\T}=\Lambda^{-1/2} E^T \bx$ with $A^{\T}=(\Lambda^{-1/2}E^T)A(E\Lambda^{1/2})$ and $C^{\T}=C(E\Lambda^{1/2})$ such that the new covariance of $\bx^{\T}$ is the identity matrix, i.e., $Q^{\T}=\mathbf{I}$. Thus one can constrain $Q=\mathbf{I}$ without loss of generality.

For Constraint 2, the components of the state vector can be arbitrarily reordered; this corresponds to swapping the columns of $C$ and $A$. Therefore,the order of the columns of matrix $C$ must be fixed. We follow Roweis and Ghahramani and choose the order by decreasing the norms of columns of $C$.

Additionally, $V_0$ is set to zero, meaning the starting state $\bx_0=\mathbf{\pi}_0$ is an unknown constant instead of a random variable, since there is only a single chain of time series in the neuroimaging application. To estimate $V_0$ accurately, multiple series of observations are required.

The following three constraints are further applied to achieve a more useful model.
\vspace*{-3mm}
\begin{equation*}\label{eqn:constraints2}
\begin{aligned}
&\textbf{Constraint 4: }R\text{ is a diagonal matrix}\\
&\textbf{Constraint 5: }A\text{ is sparse}\\
&\textbf{Constraint 6: }C\text{ has smooth columns}
\end{aligned}
\end{equation*}

Consider the case where the observed data are high dimensional and the $R$ matrix is very large. One can not accurately estimate the many free parameters in $R$ with limited observed data. Therefore some constraints on $R$ will help with inferential accuracy, by virtue of significantly reducing variance while not adding too much bias. In the simplest case, $R$ is set to an identity matrix or its multiple. More generally, one can also constrain matrix $R$ to be diagonal. In the static model with no temporal dynamics, a diagonal $R$ is equivalent to the generic Factor Analysis method, while multiples of the identity $R$ matrix lead to Principal Component Analysis (PCA) \citep{roweis1999unifying}.

The $A$ matrix is the transition matrix of the hidden states. In our application, it is a central construct of interest representing a so-called connectivity graph. In many applications, it is desirable for this graph to be sparse. In this work, an $\ell_1$ penalty term on $A$ is used to impose sparsity on the connectivity graph.

Similarly, for many applications, one wants the columns of $C$ to be smooth. For example, in neuroimaging data analysis, each column of $C$ can be a signal in the brain. Therefore having the signal spatially smooth can help extract meaningful information from the noisy neuroimaging data. In this context, an $\ell_2$ penalty term on $C$ is used to enforce smoothness.

With all those constraints, the model becomes:
\begin{equation}\label{eq:model0}
\begin{aligned}
	&\bx_{t+1}=A\bx_{t}+\bY'_t, \quad \bY'_t\sim N(\mathbf{0},\mathbf{I}),\quad \bx_0 = \mathbf{\pi}_0\\
	&\by_t=C\bx_t+\mathbf{v}_t,\qquad \mathbf{v}_t\sim N(\mathbf{0},R).
\end{aligned}
\end{equation}

where $A$ is a sparse matrix and $C$ has smooth columns.
Let $\theta =\{A,C,R,\mathbf{\pi}_0\}$ represents all unknown parameters and $P(\bX,\bY)$ be the likelihood for a generic LDS model, then combing model \ref{eq:model0} and the constraints on $A$ and $C$ lead us to an optimization problem
\begin{equation}\label{eqn:penaltylik}
\hat{\theta}=\argmin_{\substack{\theta}}\left\{-\log P_\theta(\bX,\bY)+\lambda_1\|A\|_1+\lambda_2\|C\|_2^2\right\}
\end{equation}
where $\lambda_1$ and $\lambda_2$ are tuning parameters and $\|\centerdot\|_p$ represents the $p$-norm of a vector. Equivalently, this optimization problem can be written as
\begin{equation}\label{eqn:penaltylikdual}
\begin{aligned}
&\text{minimize}&\left\{-\log P_\theta(\bX,\bY)\right\}&\\
&\text{subject to: }
& \alpha\|A\|_1+ (1-\alpha)\|C\|_2^2 \leq t \text{ for some }t; &\\
&& A\in \mathcal{A}_{d\times d},\ C \in \mathcal{C}_{p \times d}, R \in \mathcal{R}_{p\times p}, \pi_0 \in \mathcal{\pi}_{d\times 1}.&
\end{aligned}
\end{equation}
where $\alpha = \frac{\lambda_1}{\lambda_1 + \lambda_2}$. $\mathcal{A}_{d\times d}$ and $\mathcal{C}_{p \times d}$ are matrix spaces of $d\times d$ and $p \times d$ dimensional respectively. $\mathcal{R}_{p \times p}$ is the $p \times p$ diagonal matrix space and $\mathcal{\pi}_{d\times 1}$ is the $d$ dimensional vector space.
\section{Parameter Estimation}
The motivating application requires solving optimization problem \ref{eqn:penaltylik}: given only an observed sequence (or multiple sequences in some applications) of outputs $\bY=(\by_1,\ldots,\by_T)$, find the parameters $\theta=\{A,C,R,\mathbf{\pi}_0\}$ that maximize the likelihood of the observed data.

Parameter estimation for LDS has been investigated extensively by researchers from control theory, signal processing, machine learning and statistics. For example, in machine learning, exact and variational learning algorithms are developed for general Bayesian networks. In control theory, the corresponding area of study is known as system identification, which identifies parameters in continuous state models.

Specifically, one way to search for the maximum likelihood solution is through iterative techniques such as expectation maximization (EM) \citep{shumway1982approach}. The detailed EM steps for a generic LDS can be found in Zoubin and Geoffrey (1996) \citep{ghahramani1996parameter}. An alternative approach is to use subspace identification methods such as N4SID and PCA-ID to compute an asymptotically unbiased solution in closed form \citep{van1994n4sid,doretto2003dynamic}. In practice, determining an initial solution with subspace identification and then refining the solution with EM is an effective approach \citep{bootslearning}.

However, the above solutions can not be directly applied to optimization problem \ref{eqn:penaltylik} due to the introduced penalty terms. We therefore developed a novel algorithm called M-estimation for Reduced rank System IDentification (\mrsid), as detailed in the following.

By the chain rule, the likelihood in model \ref{eq:model0} is
\begin{equation*}\label{eqn:likelihood}
P(\bX,\bY)=P(\bY|\bX) P(\bX)= P(\bx_0)\prod\limits_{t=1}^{T}P(\bx_t|\bx_{t-1})\prod\limits_{t=1}^{T} P(\by_t|\bx_t)=\prod\limits_{t=1}^{T}P(\bx_t|\bx_{t-1})\prod\limits_{t=1}^{T} P(\by_t|\bx_t)\mathbbm{1}_{\mathbf{\pi}_0}(\bx_0)
\end{equation*}
where $\mathbbm{1}_{\mathbf{\pi}_0}(\bx_0)$ is the indicator function and conditional likelihoods are
\begin{equation*}\label{eqn:condlik}
\begin{aligned}
P(\by_t|\bx_t)&= (2\pi)^{-\frac{p}{2}}|R|^{-\frac{1}{2}}\  \text{exp}\left\{-\frac{1}{2}[\by_t-C\bx_t]^{\T}R^{-1}[\by_t-C\bx_t]\right\}\\
P(\bx_t|\bx_{t-1})
&=(2\pi)^{-\frac{d}{2}}\  \text{exp}\left\{-\frac{1}{2}[\bx_t-A\bx_{t-1}]^{\T}[\bx_t-A\bx_{t-1}]\right\}.
\end{aligned}
\end{equation*}

Then the log-likelihood, after dropping a constant, is just a sum of quadratic terms
\begin{equation}\label{eqn:loglik}
\begin{split}
\log  P(\bX,\bY)=&-\sum\limits_{t=1}^{T}\big(\frac{1}{2}[\by_t-C\bx_t]^{\T}R^{-1}[\by_t-C\bx_t]\big)-\frac{T}{2}\text{log}|R|\\
&-\sum\limits_{t=1}^{T}\big(\frac{1}{2}[\bx_t-A\bx_{t-1}]^{\T}[\bx_t-A\bx_{t-1}]\big)-\frac{T}{2}\text{log}|\mathbf{I}|+ \text{log}(\mathbbm{1}_{\mathbf{\pi}_0}(\bx_0)).
\end{split}
\end{equation}

Replace $\log  P(\bX,\bY)$ with equation \ref{eqn:loglik}, optimization problem \ref{eqn:penaltylik} is
\begin{equation}\label{eqn:penaltylik2}
\begin{split}
\hat{\theta}=\argmin_{\substack{\theta}}\biggl\{&\sum\limits_{t=1}^{T}\big(\frac{1}{2}[\by_t-C\bx_t]^{\T}R^{-1}[\by_t-C\bx_t]\big)-\frac{T}{2}\text{log}|R|\\
&+\sum\limits_{t=1}^{T}\big(\frac{1}{2}[\bx_t-A\bx_{t-1}]^{\T}[\bx_t-A\bx_{t-1}]\big)-\frac{T}{2}\text{log}|\mathbf{I}| - \text{log}(\mathbbm{1}_{\mathbf{\pi}_0}(\bx_0))\\
&+\lambda_1\|A\|_1+\lambda_2\|C\|_2^2\biggr\}.
\end{split}
\end{equation}

Denote the target function in the curly braces  as $\mathbf{\Phi}(\theta,\bY,\bX)$, then $\mathbf{\Phi}$ can be optimized with an Expectation-Maximization (EM) algorithm.

\subsection{E Step}
The E step of EM requires computing the expected log likelihood, $\Gamma = E[\log P(\bX,\bY|\bY)]$. This quantity depends on three expectations: $E[\bx_t|\bY]$, $E[\bx_t\bx_t^{\T}|\bY]$ and $E[\bx_t\bx_{t-1}^{\T}|\bY]$. We denote their finite sample estimators by:
\begin{equation}\label{eq:expecs}
\hat{\bx}_t \equiv E[\mathbf{x_t}|\bY],\  \hat{P}_t  \equiv E[\bx_t\bx_t^{\T}|\bY],\  \hat{P}_{t,t-1}  \equiv E[\bx_t\bx_{t-1}^{\T}|\bY].
\end{equation}

Expectations \ref{eq:expecs} are estimated with a Kalman filter/smoother, which is detailed in the Appendix on the PLDS Git Repo. Notice that all expectations are taken with respect to the current estimations of parameters.
\subsection{M Step}
The parameters are $\theta =\{A,C,R,\mathbf{\pi}_0\}$. Each of them is estimated by taking the corresponding partial derivatives of $\mathbf{\Phi}(\theta,\bY,\bx)$, setting to zero and solving.

Denote estimations from previous step as $\theta^{\text{old}} =\{A^{\text{old}},C^{\text{old}},R^{\text{old}},\mathbf{\pi}_0^{\text{old}}\}$ and current estimations as $\theta^{\text{new}} =\{A^{\text{new}},C^{\text{new}},R^{\text{new}},\mathbf{\pi}_0^{\text{new}}\}$. Estimation for output noise covariance $R$ has closed form solution,
\begin{equation}\label{eq:updateR}
\begin{aligned}
\frac{\partial \mathbf{\Phi}}{\partial R^{-1}} &= \frac{T}{2}R - \sum\limits_{t=1}^T\bigl(\frac{1}{2}\by_t\by_t^{\T} - C\hat{\bx}_t\by_t^{\T}+\frac{1}{2}C\hat{P}_tC^{\T}\bigr) =0 \\
\implies R &= \frac{1}{T}\sum\limits_{t=1}^{T}(\by_t\by_t^{\T}-C^{\text{new}}\hat{\bx}_t\by_t^{\T})\\
\implies R^{\text{new}} &= \diag \biggl\{\frac{1}{T}\sum\limits_{t=1}^{T}(\by_t\by_t^{\T}-C\hat{\bx}_t\by_t^{\T})\biggr\}
\end{aligned}
\end{equation}
In the bottom line, $\diag$ denotes to only extract the diagonal terms of the matrix $R$, as we constrain $R$ to be diagonal in Constraint 4.

Estimation for initial state also has closed form. The relevant term $\log(\mathbbm{1}_{\mathbf{\pi}_0}(\hat{\bx}_0))$ is minimized only when $\mathbf{\pi}_0^{\text{new}} = \hat{\bx}_0$.

Estimation for transition matrix $C$ also has closed form solution, and the solution can be derived by rearranging the terms properly. Terms relevant to $C$ in equation \ref{eqn:penaltylik2} are
\begin{equation}\label{eq:penaltylik1}
f_{\lambda_2}(C;\bX,\bY) = \sum\limits_{t=1}^{T}\left(\frac{1}{2}[\by_t-C\bx_t]^{\T}R^{-1}[\by_t-C\bx_t]\right)+\lambda_2 \|C\|_2.
\end{equation}

In $f_{\lambda_2}(C;\bX,\bY)$, $C$ is a matrix, we vectorized it to ease optimization and notation. Here we follow the methods of \citet{turlach2005simultaneous}. Without loss of generality, assume $R$ is the identity matrix in equation \ref{eq:penaltylik1}; otherwise, one can always write equation \ref{eq:penaltylik1} as
\begin{equation*}
\sum\limits_{t=1}^{T}\left(\frac{1}{2}[R^{-\frac{1}{2}}y_t-R^{-\frac{1}{2}}Cx_t]^{\T}[R^{-\frac{1}{2}y_t}-R^{-\frac{1}{2}}Cx_t]\right) + \lambda_2 \|R^{-\frac{1}{2}}C\|
\end{equation*}
Let $\bY' = (y_{11},\ldots,y_{T1},y_{12},\ldots,y_{T2},\ldots,y_{1p},\ldots,y_{Tp})^{\T}$
be a $Tp\times 1$ vector from rearranging  $\bY$. In addition, let
\[
\bX' = \begin{pmatrix}
\bX^{\T}&&\\
&\ddots&\\
&&\bX^{\T}
\end{pmatrix}_{pT\times pd}.
\]
Finally, vectorize $C^{\text{old}}$ as
\begin{equation}\label{eq:vectorizec}
\mathbf{c}^{\text{old}} = (C_{11}^{\text{old}},\ldots,C_{1d}^{\text{old}},C_{21}^{\text{old}},\ldots,C_{2d}^{\text{old}},C_{p1}^{\text{old}},\ldots,C_{pd}^{\text{old}})^{\T}
\end{equation}
where $C_{ij}$ is the element at row $i$ and column $j$ of $C$. With these new notations, the equation \ref{eq:penaltylik1} is equivalent to
\begin{equation}\label{eq:penaltylik11}
f_{\lambda_2}(C;\bX,\bY) = \|\bY'  - \mathbf{X' c}\|_2^2 + \lambda_2\|\mathbf{c}\|_2^2.
\end{equation}
With the Tikhonov regularization \citep{tikhonov1943stability}, equation \ref{eq:penaltylik11} has closed form solution
\begin{equation}\label{eq:updatec}
\begin{aligned}
\mathbf{c}^{\text{new}} &= (\bX'^{\T}\bX' + \lambda_2\mathbf{I})^{-1}\bX'^{\T}\bY'\\
C^{\text{new}} &=\text{Rearrange } \mathbf{c}^{\text{new}} \text{ by equation }\ref{eq:vectorizec}
\end{aligned}
\end{equation}

Now let's look at parameter $A$. Terms involving $A$ in equation \ref{eqn:penaltylik2} are,
\begin{equation}\label{eq:penaltylik2}
f_{\lambda_1}(A;\bX,\bY) = \sum\limits_{t=1}^{T}\big(\frac{1}{2}[\bx_t-A\bx_{t-1}]^{\T}[\bx_t-A\bx_{t-1}]\big)+\lambda_1 \|A\|_1.
\end{equation}

Similar to what we have done to $C$, equation \ref{eq:penaltylik2} is equivalent to
\begin{equation}\label{eq:penaltylik21}
f_{\lambda_1}(A;\bX,\bY) =  \|\mathbf{z}  - \mathbf{Za}\|_2^2 + \lambda_1\|\mathbf{a}\|_1.
\end{equation}
where $\mathbf{z}$ is a $Td \times 1$ vector from rearranging $\bX$ and $\mathbf{Z}$ is a block diagonal matrix with diagonal component $Z^{\T} =(\bx_0,\ldots,\bx_{T-1})^{\T}$. Unfortunately, equation \ref{eq:penaltylik21} does not have closed form solution due to the $\ell_1$ term.


Though not having a closed form solution, $f_{\lambda_1}(A;\bX,\bY)$ can be solved numerically with a Fast Iterative Shrinkage-Thresholding Algorithm (FISTA) \citep{beck2009fast}. FISTA is an accelerated version of the Iterative Shrinkage-Threshholding Algorithm (ISTA) \citep{daubechies2004iterative}. ISTA is linearly convergent while FISTA is quadratic convergent. Steps of a general FISTA algorithm can be found in the Appendix on the PLDS Git Repo.

FISTA requires calculating the Lipschitz constant $L$ for $\nabla\mathbf{g(z)}=\mathbf{Z}^{\T}(\mathbf{Z}\mathbf{a} -\mathbf{z})$, where $\mathbf{g}(\mathbf{z})=\|\mathbf{Z}^{\T}\mathbf{a} -\mathbf{z}\|_2^2$. Denote $\|Z\|$ as the induced norm of matrix $Z$, then $L$ is
\[
L = \sup_{\substack{x\neq y}}\frac{\|\mathbf{Z}^{\T}(\mathbf{Z}x- \mathbf{Z}y)\|}{\|x-y\|}=\sup_{\substack{x\neq 0}}\frac{\|\mathbf{Z}^{\T}\mathbf{Z}x\|}{\|x\|}\leq\|\mathbf{Z}^{\T}\|\|\mathbf{Z}\| = \|Z^{\T}\|\|Z\|.
\]

With FISTA and $L$, matrix $A$ can be updated:
\begin{equation}\label{eq:updatea}
A^{\text{new}} = \text{FISTA}(\|\mathbf{Z}^{\T}\mathbf{a}^{\text{old}} -\mathbf{z}\|_2^2,\quad \lambda_1)
\end{equation}

\subsection{Initialization}\label{sec:initial}
$R$ matrix is initialized as the identify matrix, while $\mathbf{\pi}_0$ as $\mathbf{0}$ vector. For $A$ and $C$, denote $\bY = \left[\mathbf{y_1},\cdots,\mathbf{y_T}\right]$, a $p\times T$ matrix, then the singular value decomposition (SVD) of $\bY$ is $\bY = \mathbf{UDV^{\T}} \approx \mathbf{U}_{p \times d} \mathbf{D}_{d \times d} \mathbf{V}_{d \times T}^{\T} =\mathbf{U}_{p\times d}\bX_{d \times T}$, where $\mathbf{U}_{p \times d}$ is the first $d$ columns of $\mathbf{U}$ and $\mathbf{D}_{d\times d}$ is the upper left block of $\mathbf{D}$. This notation also applies to $\mathbf{V}^{\T}_{d \times T}$.
$C$ is then initialzed as $\mathbf{U}_{p\times d}$, while the columns of $\bX_{d \times T}$ are used as input for a vector autoregressive (VAR) model to estimate the initial value for $A$.

\subsection{The Complete EM}\label{sec:em}
The complete EM algorithm for \mrsid~is addressed in Table \oldref{tab:em}. Notice that all the terms involving $\bX$ in the M-step are approximated with the conditional expectations calculated in E-step.\\
\begin{table}
\captionof{table}{The Complete EM Algorithm}
\label{tab:em}
\begin{tabular}{l}
\hline
\textbf{Algorithm } EM Algorithm for \mrsid\\
\hline
\textbf{M Step}\\
1. $R^{\text{new}}=\diag\biggl\{\frac{1}{T}\sum\limits_{t=1}^{T}(\by_t\by_t^{\T}-C^{\text{old}} \hat{\bx}_t\by_t^{\T})\biggr\}$, as in equation \ref{eq:updateR}\\
2. $\mathbf{\pi}_0^{\text{new}}=\hat{\bx}_0$\\
3. Update $C^{\text{new}}$, as in equation \ref{eq:updatec}\\
4. Update $A^{\text{new}}$ with FISTA, as in equation \ref{eq:updatea}\\
5. Stop when difference between estimations from this step and previous step\\
  $\quad$ is less than tolerance or maximum number of iterations reached.\\
\hline
\textbf{E Step}\\
0. Initialize $\theta =\{A,C,R,\mathbf{\pi}_0\}$ as in Section \oldref{sec:initial}, if first loop\\
1. Update the expectations in \ref{eq:expecs} with the Kalman filter smoother\\
\hline
\end{tabular}
\end{table}

\subsection{Improving Computational Efficiency}
The major factors that affect the efficiency and scalability of the above EM algorithm involve the storage and computations of covariance matrix $R$. The following computational techniques are utilized to make the code highly efficient and scalable.

First, a sparse matrix is used to represent R. When dimension $p$ gets higher, the size of $R$ increase quadratically, which will easily exceed the memory capacity of a computer.  Fortunately, with Constraint 4, $R$ is sparse and can be represented with a sparse matrix. For example, when $p=10,000$, the full $R$ matrix consumes over 100 gigabyte of memory, while the sparse matrix takes less than 1 megabyte.

In addition, to update $R$ in the M step, directly calculate its diagonal without calculating the full matrix $R$.

Finally, in the E-step, a term $K_t=V_t^{t-1}C^{\T}(CV_t^{t-1}C^{\T}+R)^{-1}$ involving $R$ need to be calculated, which involves the inverse of a large square matrix of dimension p by p. As stated previously, such a matrix exceeds available memory when p is high. The Woodbury Matrix Identity is employed to turn a high dimensional inverse to low dimensional problem: $(CV_t^{t-1}C^{\T}+R)^{-1} = R^{-1} - R^{-1}C[(V_t^{t-1})^{-1} + C^{\T}R^{-1}C]^{-1}C^{\T}R^{-1}$.

Also note that quantities like $R^{-1}$ and $C^{\T}R^{-1}C$ can be pre-computed and reused throughout the E step. With the above three techniques, the EM algorithm can scale to very high dimensions in terms of $p$, $d$ and $T$, without causing any computational issues.

\subsection{The Data}

The Kirby 21 data are resting-state fMRI scans consisting of a test-retest dataset previously acquired at the FM Kirby Research Center at the Kennedy Krieger Institute, Johns Hopkins University \citep{landman2011multi}. Twenty-one healthy volunteers with no history of neurological disease each underwent two separate resting state fMRI sessions on the same scanner: a 3T MR scanner utilizing a body coil with a 2D echoplanar (EPI) sequence and eight channel phased array SENSitivity Encoding (SENSE; factor of 2) with the following parameters: TR 2s; 3mm$\times$3mm in plane resolution; slice gap 1mm; and total imaging time of 7 minutes and 14 seconds.

The Human Connectome Project is a systematic effort to map macroscopic human brain circuits and their relationship to behavior in a large population of healthy adults \citep{van2013wu,moeller2010multiband,feinberg2010multiplexed}. MR scanning includes four imaging modalities, acquired at high resolutions: structural MRI, resting-state fMRI (rfMRI), task fMRI (tfMRI), and diffusion MRI (dMRI). All 1200 subjects are scanned using all four of these modalities on a customized 3 T scanner.  All scans consist of 1200 time points.

\section{Results}
\subsection{Parameter Estimation}
\label{sec:lowdsim}
Two simulations of different dimensions are performed to demonstrate the model and its parameter estimations.

In the low dimensional setting, $p = 300$, $d = 10$ and $T = 100$. A is first generated from a random matrix, then elements with small absolute values are then truncated such that 20 percent of elements are zeros. After that a multiple of the identity matrix is added to A. $A$ is then scaled to make sure its eigenvalues fall within $[-1,1]$ to avoid diverging time series. Matrix C is generated as follows. Each column contains random samples from a standard Gaussian distribution. Then each column is sorted in ascending order. Covariance $Q$ is the identity matrix and covariance $R$ is a multiple of the identity matrix. At time 0, a zero vector $\mathbf{0}$ is used as the value of $\mathbf{x_0}$. Pseudocode for data generation can be found in the Appendix on PLDS Git Repo.

In the high-dimensional setting, $p = 10000$, $d = 30$ and $T = 100$. The parameter are generated in a the same manner.

To evaluate the accuracy of estimations, we elect to define the distance between two matrices $A$ and $B$ is defined as follows
\begin{equation}\label{eq:distance}
d(A,B) = \argmin_{P\in P(n)}\left\{\log\bigl[\frac{n}{\text{\footnotesize Trace}(P\times C_{A,B})}\bigr]\right\}.
\end{equation}
where $C_{A,B}$ is the correlation matrix between columns of A and B, $P(n)$ is the collection of all the permutation matrices of order n and $P$ is a permutation matrix.

As a result of the way it's defined, $d(A,B)$ is invariant to the scales of columns of $A$ and $B$. It is also invariant to a permutation of columns of either matrix. The calculation of $d(A,B)$ is exactly a linear assignment problem and can be solved in polynomial time with the Hungarian algorithm \citep{kuhn1955hungarian}.

Both the generic LDS and the penalized LDS are applied to the simulation data. As the true parameters are sparse, we expect that the penalized algorithms would yield better estimations with appropriately chosen penalty parameters. When the penalties are approaching 0, the penalized algorithm should converge to the generic model. In addition, when the penalties are too big or too small compared to the optimal values, estimations might be less accurate.

A sequence of tuning parameters $\lambda_C$ are utilized, ranging from $10^{-6}$ to $10^4$. $\lambda_A = k \lambda_C$ is set to increase proportionally with $\lambda_C$, where $k$ is a constant.

\begin{figure}
\centering
\subfigure[Low dimensional setting]{%
\includegraphics[scale=.43]{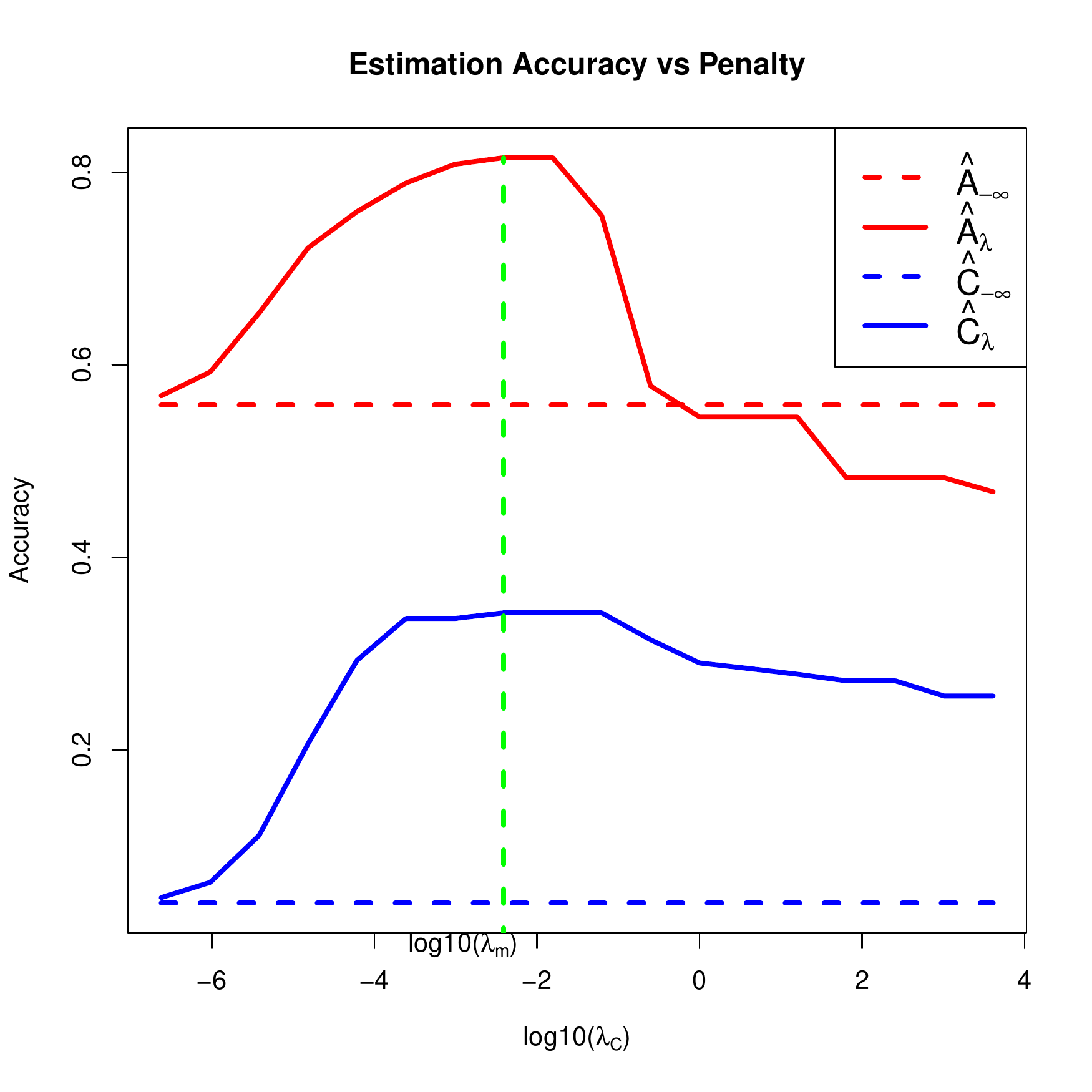}
}
\subfigure[High dimensional setting]{%
\includegraphics[scale=.43]{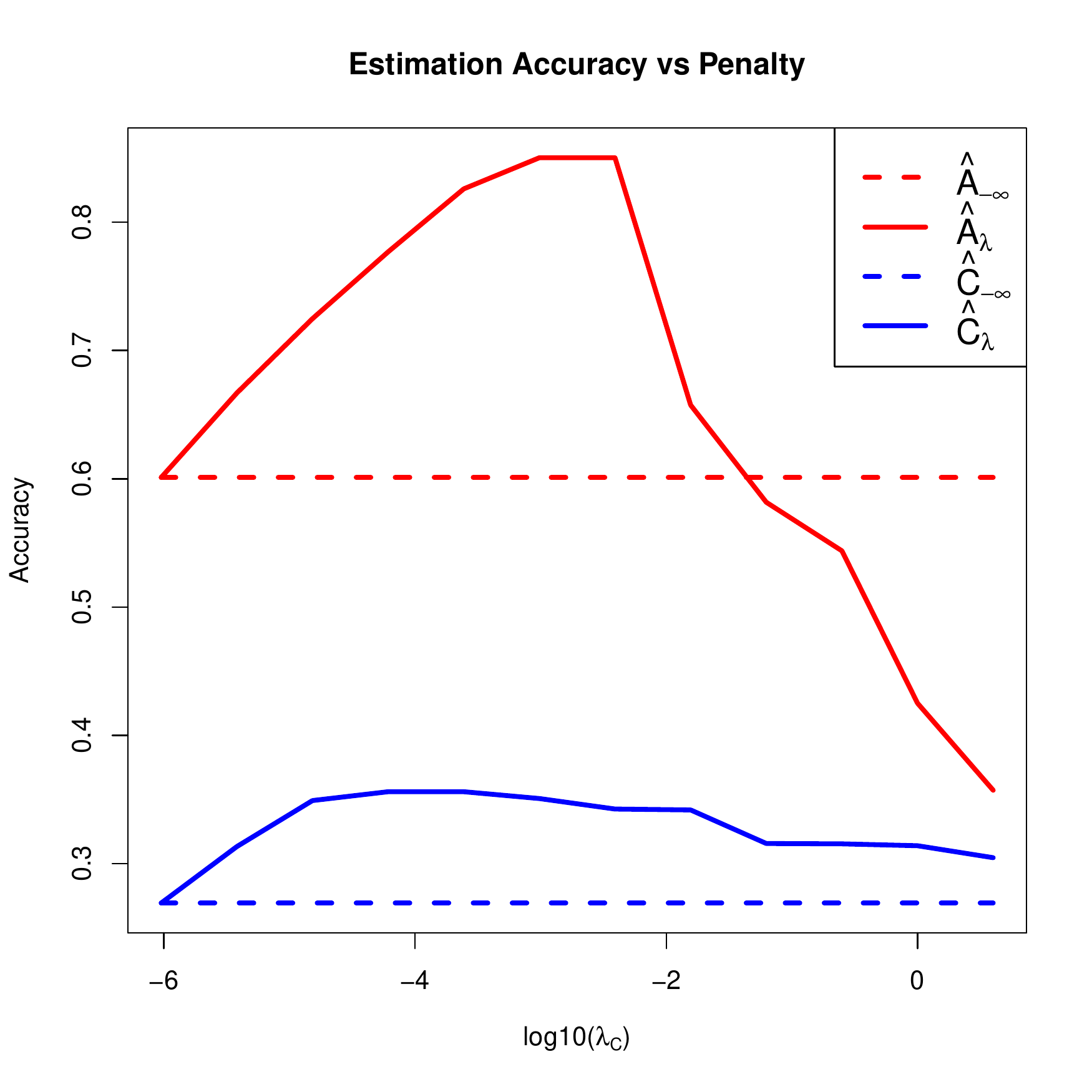}
}
\caption{x axis is tuning parameter $\lambda_C$ under log scale and y axis is the distance between truth and estimations; $\lambda_A$ is increasing proportionally with $\lambda_C$. One can see that in both the low dimensional and hight dimensional setting, estimation accuracies for $A$ and $C$ first increase then decrease as penalty increases.}
\label{fig:low-high-d-sim}
\end{figure}

Estimation accuracies are plotted against penalty size $\lambda_C$ in Figure \oldref{fig:low-high-d-sim}. Results from LDS and \mrsid~are overlayed in one plot for comparison. As the figure shows, \mrsid~converges to the LDS when the penalties are approaching zero. Estimation accuracies first increase with penalty size and then decrease due to over-shrinkage.

\begin{figure}
 \centering
 \includegraphics[scale=.6]{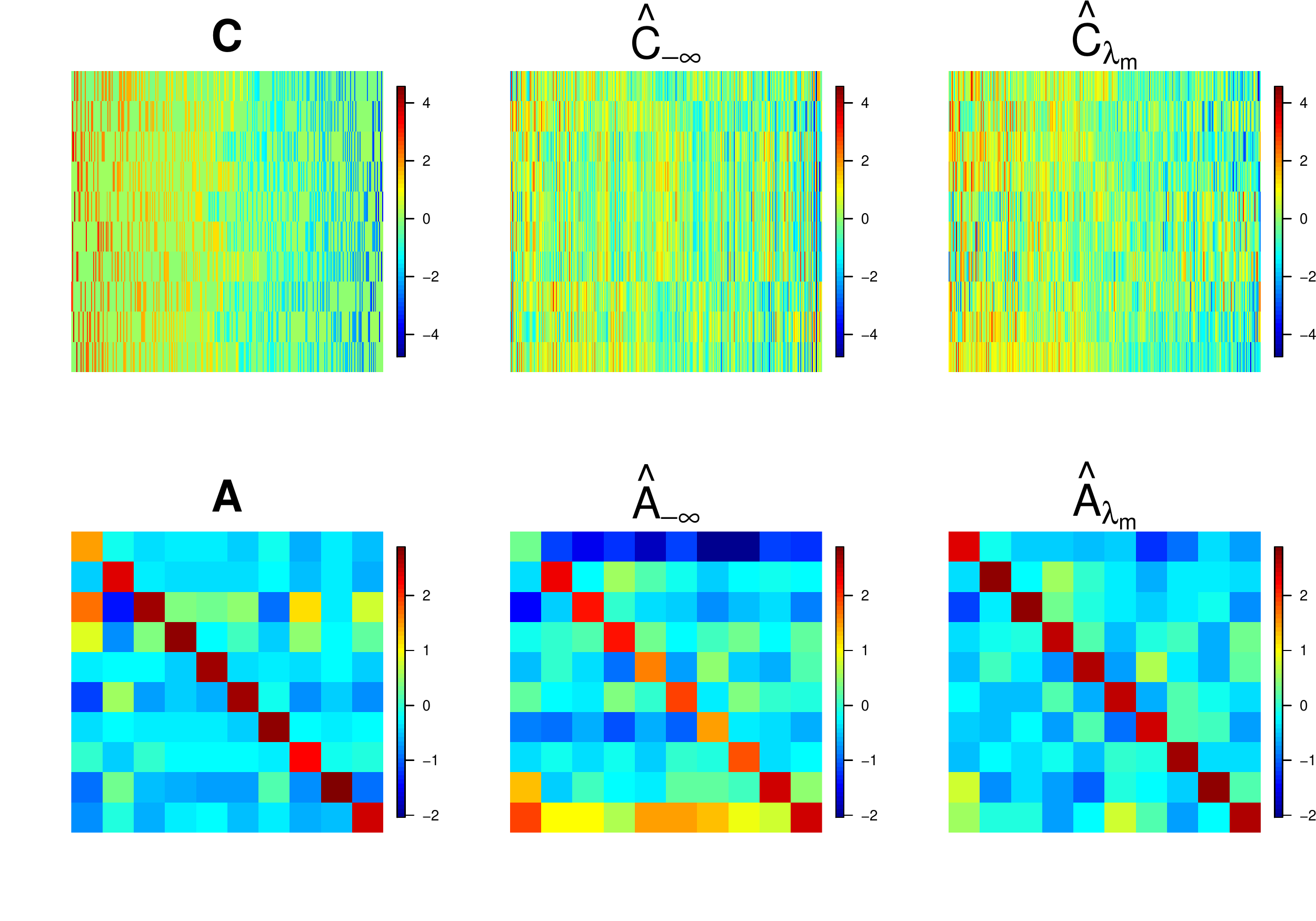}
 \captionof{figure}{Row 1: A truth; non-penalized estimation of A; optimally penalized estimation of A. Row 2: C truth; non-penalized estimation of C; optimally penalized estimation of C.}
 \label{fig:heatmap}
\end{figure}

As a concrete example, estimations from both methods are compared to the true values of parameters in Figure \oldref{fig:heatmap}. One can see that true values in each column of $C$ matrix are decreasing smoothly. $\hat{C}_{\lambda_m}$, which is estimated with optimal penalties $\lambda_C = \lambda_m$ and $\lambda_A = k\lambda_m$, shows similar pattern. In terms of $A$, the true value is sparse with many $0$ (blue) values. \mrsid~estimation $\hat{A}_{\lambda_m}$ is also sparse, denoted by the off-diagonal 0 (blue) values. However, LDS estimation $\hat{A}_{\lambda_{-\infty}}$ is not sparse, with many positive (yellow and red) off-diagonal values.

In addition to the improved estimation accuracy, the proposed algorithm is also computational efficiency and highly scalable. As a demonstration, we measure the running times of multiple simulation scenarios and summarize them in Table \oldref{tab:runningTime}. When both $p$ and $d$ are high dimensional, the algorithm can still solve the problem in a reasonable time.
\begin{table}
\centering
\captionof{table}{\mrsid~Running Time}
\label{tab:runningTime}
\begin{tabular}{c|ccccc}
\hline\hline
$p$ & 100 & 1000 & 10000 & 100000 & 100000\\
\hline
$d$ & 10 & 30 & 50 & 100 & 500 \\
\hline
$T$ & 100 & 300 & 500 & 1000 & 1000 \\
\hline
Time (min)& 0.07 & 0.30  & 5.15 & 111.38 & 1127.18 \\
\hline\hline
\end{tabular}
\end{table}
\subsection{Making Predictions}
\begin{figure}
\centering
\includegraphics[scale=0.46]{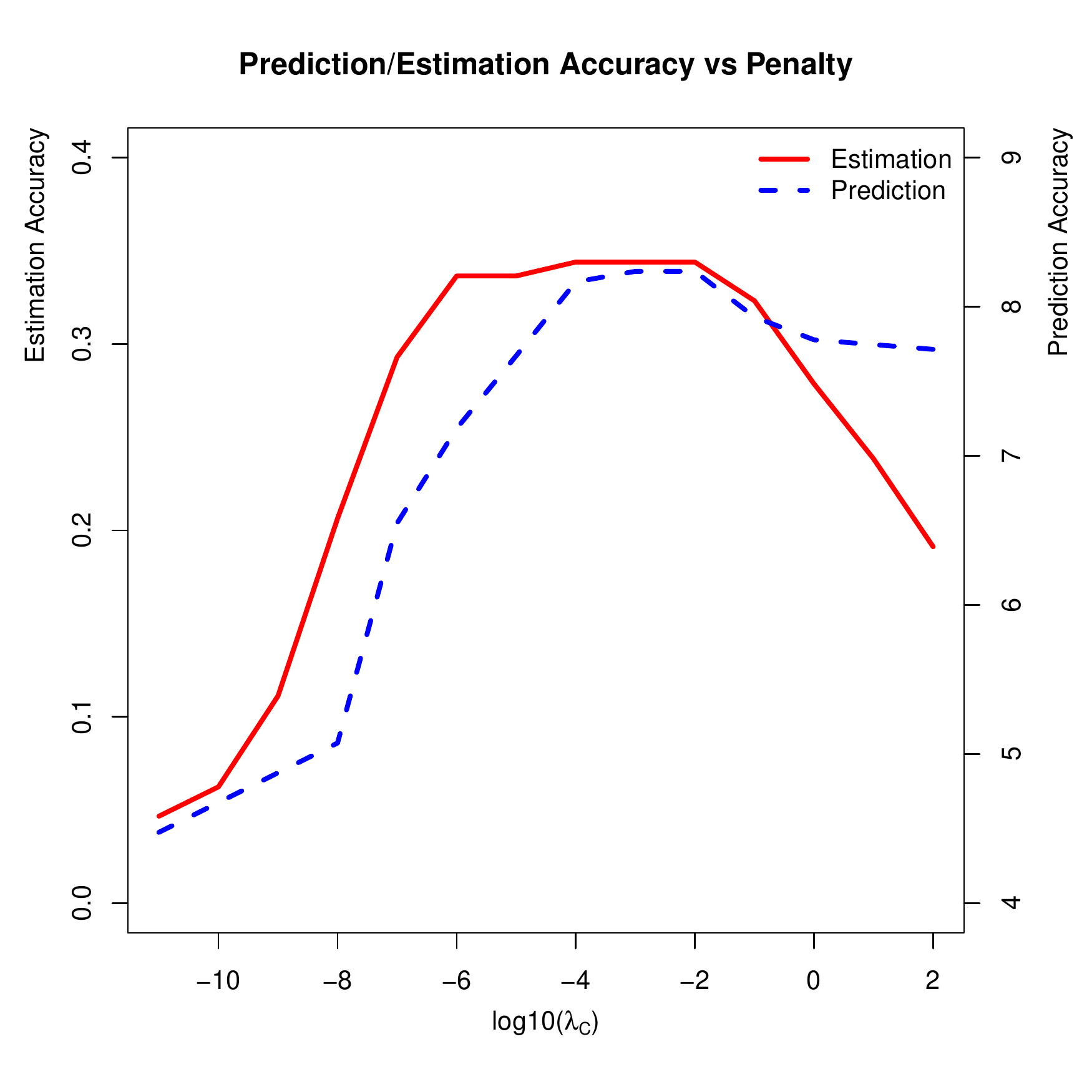}
\captionof{figure}{Estimation and prediction accuracies. The x-axis is the penalty size under log scale, while y-axis is the estimation and prediction accuracies. One can see that the penalty that yields the most accurate estimation also gives the best predictions.}
\label{fig:estpredaccuracy}
\end{figure}

Another perspective when considering \mrsid~model is its ability to make predictions. When the parameters $\mathbf{\theta}$ and the latent states $x_T$ are estimated, one can first use estimated $x_T$ to predict $x_{T+1}$ and use $x_{T+1}$ to predict $y_{T+1}$. Similarly, more predictions $y_{T+2},\ldots, y_{T+k}$ can be made. Intuitively, properly chosen penalties give better estimations and good estimations should give more accurate predictions. This idea is demonstrated with a simulation. The parameter settings for this simulation follow Section \oldref{sec:lowdsim}. The correlation between the predicted signal and true signal is used as a measure of prediction accuracy. The prediction accuracy over penalty size is shown in Figure \oldref{fig:estpredaccuracy}.

From the plots one can see that the prediction accuracy first improves then drops when the penalties increase. The prediction accuracy peaks when the penalty coefficient $\lambda_A$ and $\lambda_C$ are around $10^{-3}$. This make sense as the same $\lambda$ pair also gives the best estimation for coefficients $A$ and $C$, as in Figure \oldref{fig:low-high-d-sim}. This latter observation provides us a way to pick tuning parameters in real applications, as detailed in Section \oldref{sec:application}.

\section{Application}
\label{sec:application}
When applied to fMRI data analysis, the model has very good interpretability. Each $\by_t$ is a time step including the entire brain volume. Each column of the $C$ matrix is interpreted as a time-invariant brain ``point spread function''. At each time point, the observed brain image, $\by_t$, is a linear mixture of latent co-assemblies of neural activity  $\bx_t$. Matrix $A$ describes how $\bx_t$ evolves over time. $A$ can also be viewed as a directed graph if each neural assembly is treated as a vertex. Each neural assembly is spatially smooth and connectivities across them are empirically sparse. This naturally fits into the sparsity and smoothness assumptions in \mrsid.

\mrsid~is applied to analyze the motor cortex of human brains from the KIRBY 21 Data. In this application, test-retest scans from two subjects are analyzed. The imaging data are first preprocessed with FSL, a comprehensive library of analysis tools for fMRI, MRI and DTI brain imaging data \citep{smith2004advances}. FSL is used for spatial smoothing with Gaussian kernel. Then \mrsid~is applied on the smoothed data.

The following are basic descriptions of the data and model parameters: number of voxels, $p = 7396$; Number of scans, $T = 210$; Number of latent states, $d = 11$. Tuning parameters: $\lambda_A = \lambda_C = 10^{-5}$. Different combinations of $\lambda_A$ and $\lambda_C$, where $\lambda_A = \lambda_C$ is applied to the data. The values range from $10^{-10}$ to $10^{4}$. Then the estimations are used to make predictions. The combination that gives the best predictions is used here. One can also use a grid of combinations, but it is time consuming. Max number of iterations for EM and the regularized subproblems are both 30 steps.


A flexible method to choose the number of latent states involves the profile likelihood method proposed by Zhu et al. \citep{zhu2006automatic}. The method assumes eigenvalues of the data matrix come from a mixed Gaussian and use profile likelihood to pick the optimal number of latent states. Apply the method to all four scans, the numbers of latents states selected are 11, 6, 14 and 15 respectively. Their average, $d=11$, is used.

Denote the $A$ matrix estimation for the second scan of subject one as $A_{12}$. Similar notations apply to the other scans. Then the similarities among the four matrices are summarized in Table \oldref{tab:similarity}. The distance measure in Equation \ref{eq:distance} is used. Another permutation invariant measure of distance between two square matrices, the Amari error \citep{amari1996new}, is also provided in the table. The Amari error between $A$ and $\hat{A}$: $E(A,\hat{A}) = \sum\limits_{i=1}^n(\sum\limits_{j=1}^n\frac{|p_{ij}|}{\max_k |p_{ik}|}-1) + \sum\limits_{j=1}^n(\sum\limits_{i=1}^n\frac{|p_{ij}|}{\max_k|p_{kj}|}-1)$, where $P =(p_{ij})=A^{-1}\hat{A}$. Notice a smaller $d(A,B)$ or Amari error means more similarity. Among the six pairs from $A_{11},A_{12},A_{21}$ and $A_{22}$, it is expected that the pairs $(A_{11},A_{12})$ and $(A_{21},A_{22})$ give the smallest distances, as each pair comes from two scans of the same subject. This idea is validated by Table \oldref{tab:similarity}. A direct application of this result is to correctly cluster the four scans into two group, each group corresponding to a subject. This implies that the A matrices contains subject-specific information. The similarities among $A$ matrices are also shown in Figure \oldref{fig:matsim} as a heatmap. 

\begin{table}
\centering
\captionof{table}{Similarities Among Estimated $A$ Matrices}
\label{tab:similarity}
\begin{tabular}{c|cccc}
\hline
$d(\cdot,\cdot)$(Amari Error) & $A_{11}$&$A_{12}$ & $A_{21}$&$A_{22}$ \\
\hline
$A_{11}$ & $0$ &  &  &\\
$A_{12}$ & $\mathbf{0.076(0.88)}$& $0$ & &\\
$A_{21}$ & $0.105(1.05)$ & $0.095(1.08)$  & $0$ &\\
$A_{22}$ & $0.095(1.02)$ & $0.095(1.09)$ & $\mathbf{0.085(0.98)}$ & $0$ \\
\hline
\end{tabular}
\end{table}

\begin{center}
\includegraphics[scale=.4]{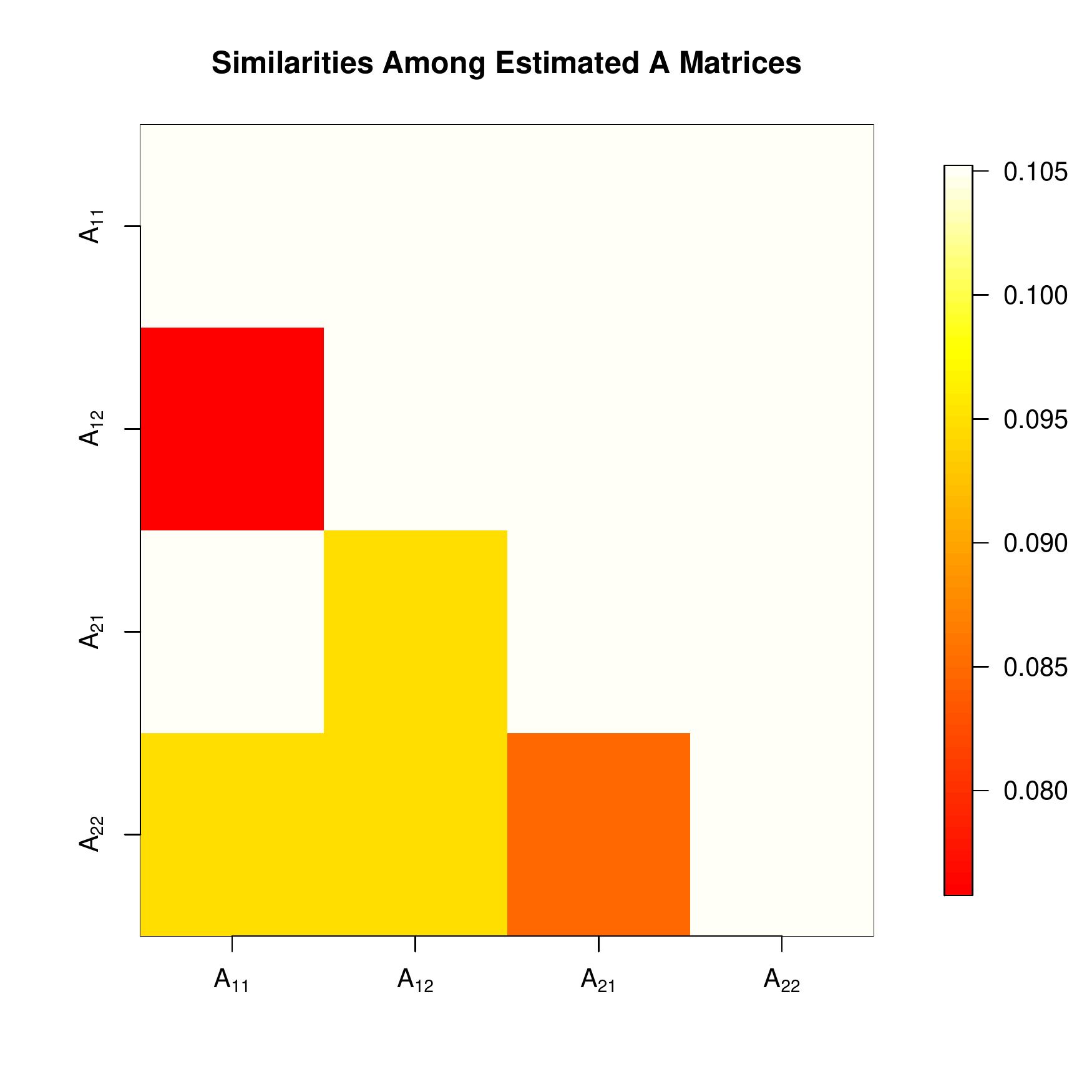}
\captionof{figure}{Similarities among the four estimated $A$ matrices. The distance $d(\cdot,\cdot)$ is used in this figure. As one can see, the two red/orange off-diagonal pixels has the minimum distances, which correspond to the pairs of $(A_{11},A_{12})$ and $(A_{21},A_{22})$ respectively. With this similarity map, one can tell which two scans are from the same subject.}
\label{fig:matsim}
\end{center}

In addition, the 3D renderings of the columns of matrix $C$ from the first scan of subject one are shown in Figure \oldref{fig:3d} (after thresholding).
It is helpful to compare those regions to other existing parcellations of the mortor cortex. As an example, the blue region in Figure \oldref{fig:3d} accurately matches the dorselmedical (DM) parcel of the five-region parcellation proposed by Nebel MB et al. \citep{nebel2014disruption}.
\begin{center}
\[
\begin{array}{lll}
\includegraphics[scale = 0.36]{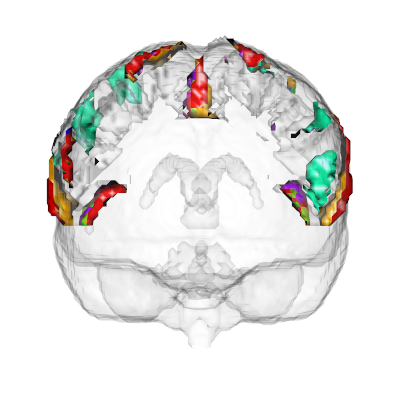} & \includegraphics[scale = 0.33]{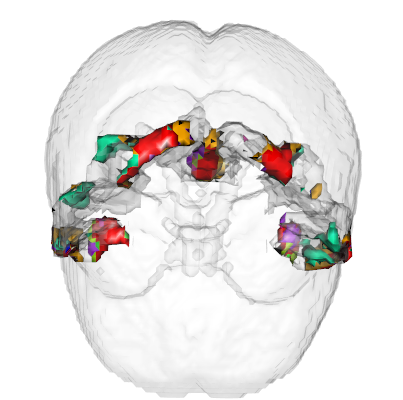} & \includegraphics[scale = 0.33]{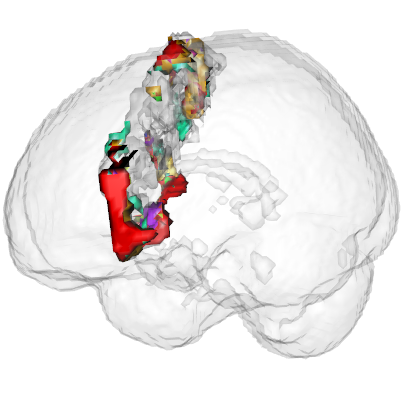}
\end{array}
\]
\captionof{figure}{3D rendering of columns of matrix $C$: estimation from the first scan of subject one shown in this plot.}
\label{fig:3d}
\end{center}

Another application of the algorithm is predicting brain signals. To demonstrate this, the algorithm is applied to the Human Connectome Project (HCP) data.
Using the profile likelihood method, $d=149$ is picked. The data has $T=1200$ time points. The first $N = 1000$ are picked as training data, while the rest are used as test data. Then both the SVD method in Section \oldref{sec:initial} and \mrsid~algorithm are used for estimations. Then $k$-step ahead predictions are made with equations \ref{eq:model0} and estimations from both methods. Pseudocode for $k$-step ahead predictions is given in Appendix on PLDS Git Repo. The prediction accuracies are shown in Figure \oldref{fig:predaccy} (left panel). One can see \mrsid~algorithm is giving significantly better predictions for the first 150 predictions compared to the SVD method. As the SVD method is also used to intialize \mrsid~algorithm, this shows that \mrsid~algorithm improves estimations from the SVD method in terms of short-term predictions. Another observation is, \mrsid~algorithm's performance get worse when one predicts into the ``long" future ($>150$ steps). This is reasonable because the prediction errors from each step will accumulate and yields deteriorating predictions as the number of steps increase.

\begin{center}
\includegraphics[scale=0.45]{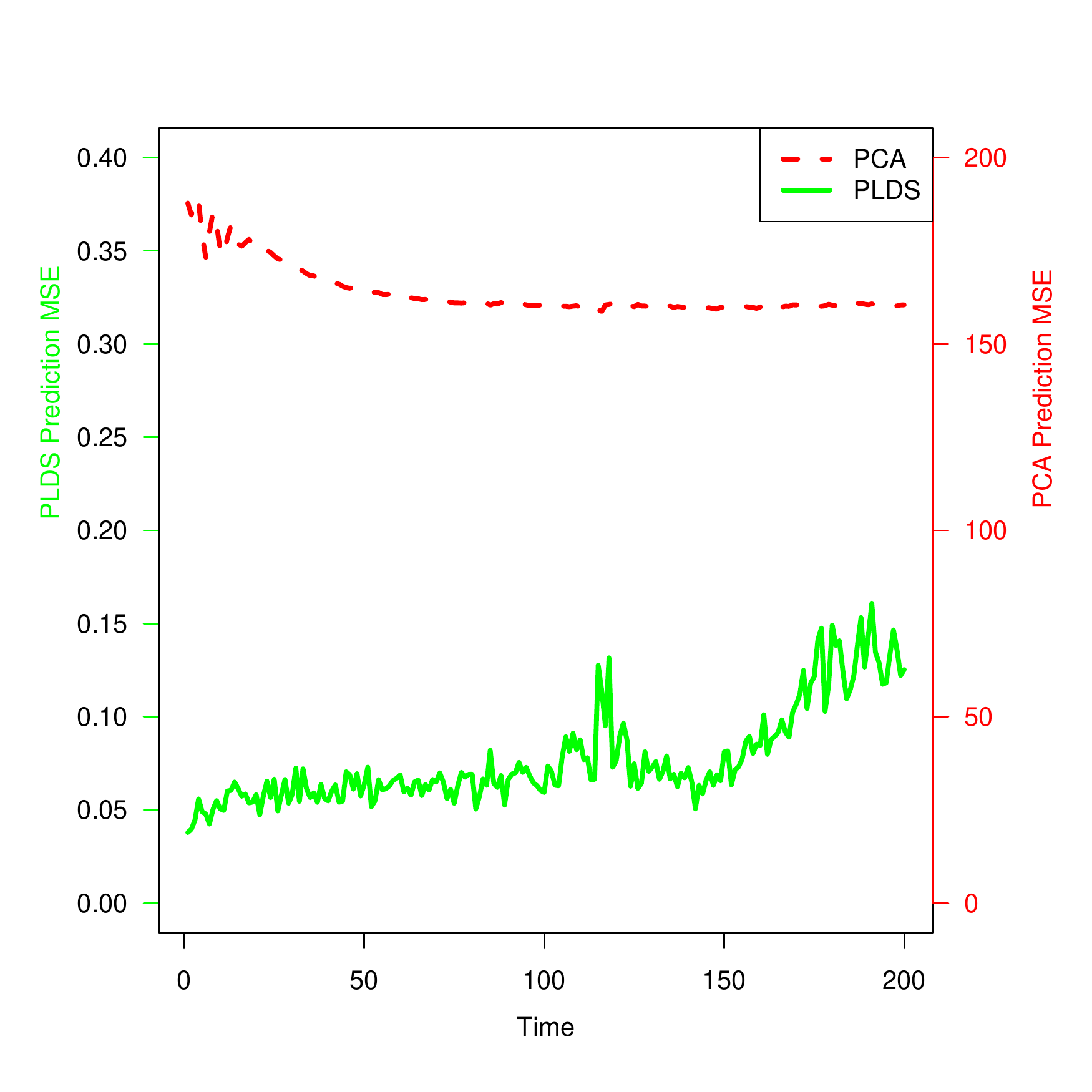}
\includegraphics[scale=0.45]{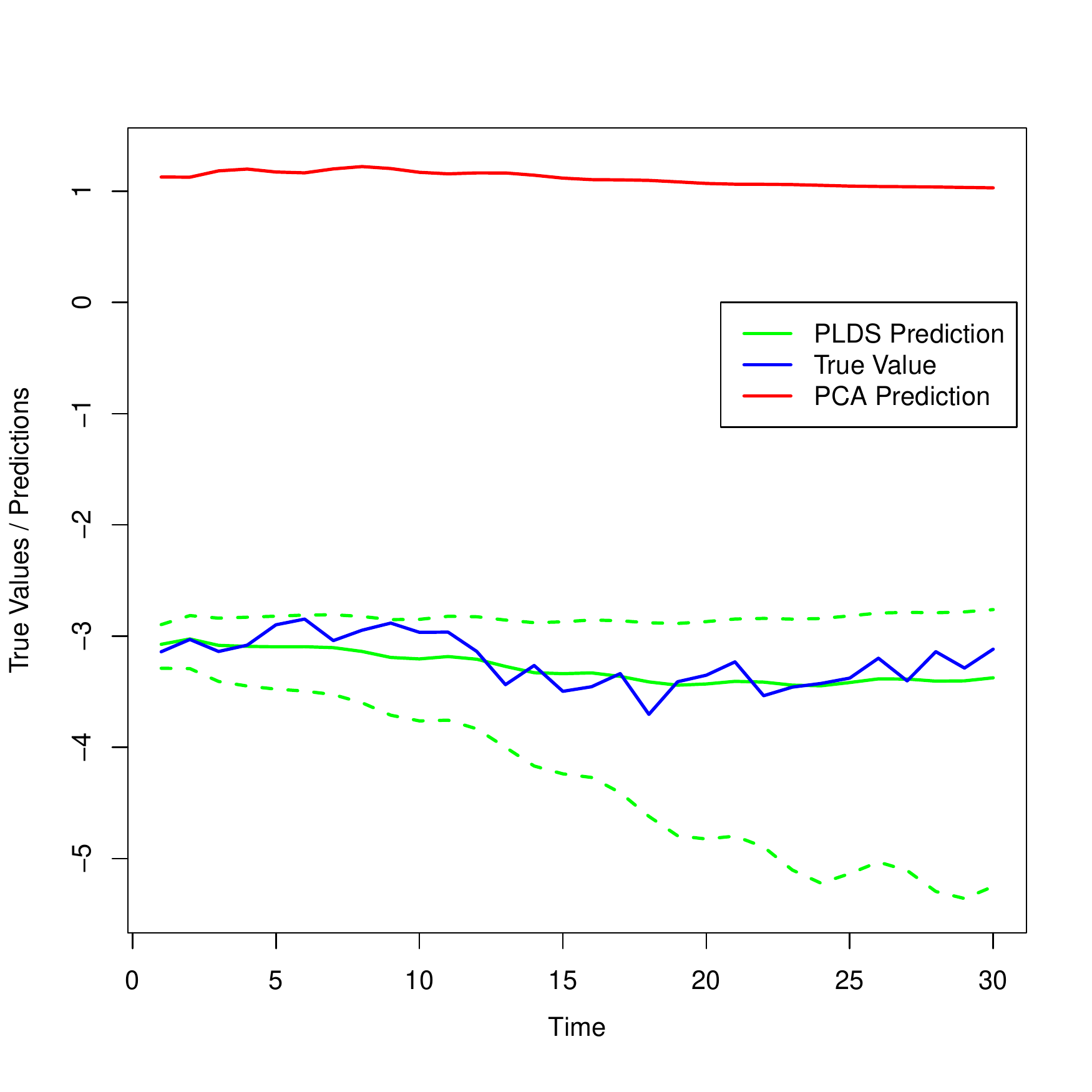}
\captionof{figure}{Prediction accuracies comparison on HCP data.
\emph{Left:} The mean squared error (MSE) is used as the accuracy measure.
\emph{Right:} Sample time series plot. The dotted green curve stands for the $60\%$ confidence band given by \mrsid~model. The true time series is averaged signals from a subsample of voxels. The predictions are also averaged over the same subsample. The confidence band is estimated based on the covariance matrix of these voxels. A subsample of 20 voxels are picked in this experiment to avoid big covariance matrices calculation. All values are log-scaled for plotting purpose.
}
\label{fig:predaccy}
\end{center}

A sample plot of the true time series and predicted values are shown in Figure \oldref{fig:predaccy} (right panel). We see that \mrsid~is giving more accurate predictions and the true signal lies in the confidence band giving by \mrsid~model. Another observation is that the confidence band is getting wider as we predict into the future, which is a result of the accumulated errors.

\section{Discussion}

By applying the proposed model to fMRI scans of the motor cortex of healthy adults, we identify limited sub-regions (networks) from the motor cortex. A statistical procedure should be further developed to match these regions to existing parcellations of the motor cortex.

In the future, this work could be extended in two important directions. First, assumptions on the covariance structures in the observation equation could be generalized. Prior knowledge could be incorporated to covariance $R$ \citep{allen2014generalized}. The general rule is that $R$ should be general enough to be flexible while sufficiently restricted to make the model useful. A lot of other platforms such as tridiagnol and upper triangular could also be considered. Mohammad et al. have discussed the impact of auto correlation on functional connectivity, which also provides us a direction for extension \citep{arbabshirani2014impact}.

Finally, the work can also be extended on the application side. Currently, only data from a single subject is analyzed. As a next step, the model can be extended to a group version and be used to analyze more subjects. The coefficients from the algorithm could be used to measure the reproducibility of the scans.

\section*{Appendix 1}
\label{sec:appendix1}

\begin{tabular}{l}
\hline
\textbf{Algorithm } Standard Kalman Filter Smoother for estimating the moments \\
$\qquad\quad\quad$ required in the E-step of an EM algorithm for a linear dynamical system\\
\hline
0. Define $\bx_t^{\tau}$ = E($\bx_t|\bY_1^{\tau}$),$\mathbf{V}_t^{\tau}=\text{Var}(\bx_t|\bY_1^{\tau})$, $\hat{\bx}_t \equiv \bx_t^T$ and $\hat{P}_t\equiv V_t^T+\bx_t^T{\bx_t^T}^{\T}$\\
1. Forward Recursions:\\
\hspace{4 mm} $\bx_t^{t-1}=A\bx_{t-1}^{t-1}$\\
\hspace{4 mm} $\mathbf{V}_t^{t-1}=A\mathbf{V}_{t-1}^{t-1}+\mathbf{Q}$\\
\hspace{4 mm} $K_t=\mathbf{V}_t^{t-1}C^{\T}(CV_t^{t-1}C^{\T}+R)^{-1}$\\
\hspace{4 mm} $\bx_t^t$ = $\bx_t^{t-1} + K_t (\by_t - C\bx_t^{t-1})$\\
\hspace{4 mm} $V_t^t=V_t^{t-1}-K_tCV_t^{t-1}$\\
\hspace{4 mm} $\bx_1^0=\mathbf{\pi}_0$, $V_1^0=\mathbf{V}_0$\\
2. Backward Recursions:\\
\hspace{4 mm} $J_{t-1} = V_{t-1}^{t-1}A^{\T}(V_t^{t-1})^{-1}$\\
\hspace{4 mm} $\bx_{t-1}^T=\bx_{t-1}^{t-1}+J_{t-1}(\mathbf{x_t^T-A\bx_{t-1}^{t-1}})$\\
\hspace{4 mm} $V_{t-1}^T = V_{t-1}^{t-1}+J_{t-1}(V_t^T-V_t^{t-1})J_{t-1}^{\T}$\\
\hspace{4 mm} $\hat{P}_{t,t-1}\equiv V_{t,t-1}^T+\bx_t^T{\bx_t^T}^{\T}$\\
\hspace{4 mm} $V_{T,T-1}^T=(I-K_TC)AV_{T-1}^{T-1}$\\
\hline
\end{tabular}

\newpage

\section*{Appendix 2}
\label{sec:appendix2}
In general, FISTA optimize a target function
\begin{equation}\label{eqn: fistatarget}
\min_{\substack{x\in \mathcal{X}}}\quad \mathbf{F(x;\lambda)} = \mathbf{g(x)}+ \mathbf{\lambda \|x\|_1}
\end{equation}
where $\mathbf{g}: R^n \rightarrow R $ is a continuously differentiable convex function and $\lambda > 0$ is the regularization parameter.

A FISTA algorithm with constant step is detailed below\\

\begin{tabular}{l}
\hline
\textbf{Algorithm } FISTA$(\mathbf{g},\lambda)$.\\
\hline
 1. Input an initial guess $\mathbf{x_0}$ and Lipschitz constant $\mathbf{L}$ for $\mathbf{\nabla g}$, set $\mathbf{y_1} = \mathbf{x_0},t_1 = 1$\\
 2. Choose $\tau \in (0,1/\mathbf{L}]$.\\
 3. Set $k \leftarrow$ 0.\\
 4. \textbf{loop}\\
 5. \hspace{10mm}		Evaluate $\mathbf{\nabla g(y_k)}$\\
 6.	\hspace{10mm}	Compute $\mathbf{x_{1}}$= $\mathbf{S_{\tau\lambda}(y_k - \tau\nabla g(y_k))}$\\
 7.	\hspace{10mm}	Compute $t_{k+1} = \frac{1+\sqrt{1 + 4 t_k^2}}{2}$\\
 8.	\hspace{10mm}	$\mathbf{y_{k+1}} = \mathbf{x_k} + \left(\frac{t_k - 1}{t_{k+1}})\right (\mathbf{x_k}-\mathbf{x_{k-1}})$\\
 9.	\hspace{10mm}	Set $k \leftarrow k+1$ \\
 10. \textbf{end loop}\\
\hline
\end{tabular}

\vspace*{10mm}
In the above
\[
\mathbf{S_\lambda (y) = (|y|-\lambda)_{+}\textbf{sign}(y)}=\left\{
\begin{array}{l l }
 y - \lambda & \text{if   } y > \lambda\\
 y + \lambda & \text{if   } y < -\lambda\\
 0 & \text{if   } |y| \leq \lambda .
\end{array}
\right.
\]

\section*{Appendix 3}
\label{sec:appendix3}
\begin{tabular}{l}
\hline
\textbf{Algorithm } $k$-step predictions with PCA and \mrsid\\
\hline
 1. Denote the estimated parameters with PCA and \mrsid~as $A_{pca},C_{pca},A_{plds},$ and $C_{plds}$.\\
 2. PCA Estimated latent states at $t=1000$: $x_{1000,pca} = $ column 1000 of $\bX_{d\times T}$ from Section \oldref{sec:initial} \\
 3. \mrsid~Estimated latent states at $t=1000$: $x_{1000,pls}$ is from E step  in Section \oldref{sec:em}\\
 4. \textbf{for i = 1 to k}\\
 5. \hspace{10mm}		$x_{1000+k,pca}\ =\ A_{pca}\ x_{999+k,pca}$\\
 6.	\hspace{10mm}	$y_{1000+k,pca}\ =\ C_{pca}\ x_{1000+k,pca}$\\
 7.	\hspace{10mm}	$x_{1000+k,plds}\ =\ A_{plds}\ x_{999+k,plds}$\\
 8.	\hspace{10mm}	$y_{1000+k,plds}\ =\ C_{plds}\ x_{1000+k,plds}$\\
 9. \textbf{end}\\
\hline
\end{tabular}

\section*{Appendix 4}
\label{sec:appendix4}
\begin{tabular}{l}
\hline
\textbf{Algorithm } Simulation Data Generation\\
\hline
 1. Denote the dimensions as $p$, $d$ and $T$ respectively\\
 2. Generate a $p\times d$ matrix $C_0$ from a standard Gaussian distribution\\
 3. Sort each column of $C_0$ in ascending order to get matrix $C$ \\
 4. Generate a $d\times d$ matrix $A_0$ from a standard Gaussian distribution\\
 5. Add a multiple of the identity matrix to $A_0$\\
 6.	Replace entries in $A_0$ with small absolute values with $0$\\
 7.	Scale $A_0$ to make sure its eigen values are between $-1$ and $1$; use $A_0$ as the A matrix\\
 8.	Let $R$ be a diagonal matrix with positive diagonal entries and $Q$ be the identity matrix \\
 9. Generate simulation data with $A, C, Q$ and $R$ \\
 10. \textbf{end}\\
\hline
\end{tabular}

\bibliographystyle{Chicago}
\bibliography{reference}
\end{document}